# Seismic-ionospheric disturbances in ionospheric TEC and plasma parameters associated with the 14 July 2019 $M_w 7.2$ Laiwui earthquake detected by the GPS and CSES

**Yuanzheng Wen**[1,2], **Guangxue Wang**[1,2], **Dan Tao**[1,2], **Zeren Zhima**[3], **Yijia Zong**[1,2], **Xuhui Shen**[3]

1. Key Laboratory of Earth Exploration and Information Techniques of Ministry of Education, Chengdu University of Technology, Chengdu, China

2. Department of Geophysics and Space Sciences, School of Geophysics, Chengdu University of Technology, Chengdu, China

3. Institute of Crustal Dynamics, China Earthquake Administration, Beijing, China

Correspondence to: Dan Tao (dan.tao@cdut.edu.cn)

**Abstract:** In this study, with cross-valid analysis of total electron content (TEC) data of the global ionospheric map (GIM) from GPS and plasma parameters data recorded by China Seismo-Electromagnetic Satellite (CSES), signatures of seismic-ionospheric perturbations related to the 14 July 2019 $M_w 7.2$ Laiwui earthquake were detected. After distinguishing the solar and geomagnetic activities, three positive temporal anomalies were found around the epicenter 1 day, 3 days and 8 days before the earthquake (14 July 2019) along with a negative anomaly 6 days after the earthquake, which also agrees well with the TEC spatial variations in latitude-longitude-time (LLT) maps. To further confirm the anomalies, the ionospheric plasma parameters (electron, $O^+$ and $He^+$ densities) recorded by the Langmuir probe (LAP) and Plasma Analyzer Package (PAP) onboard CSES were analyzed by using the moving mean method (MMM), which also presented remarkable enhancements along the orbits around the epicenter on day 2, day 4 and day 7 before the earthquake. To make the investigations more convincing, the disturbed orbits were compared with their corresponding four nearest revisiting orbits, whose results indeed indicate the existence of plasma parameters anomalies associated with the Laiwui earthquake. All these results illustrated that the GPS and CSES observed unusual ionospheric perturbations are highly associated with the $M_W 7.2$ Laiwui earthquake, which also strongly indicates the existence of pre-seismic ionospheric anomalies over the earthquake region.

## 1. Introduction

Electromagnetic phenomena possibly associated with natural disasters (earthquake, tsunami and volcanic activities) have been extensively investigated in recent years, and seismic related anomalies are the most important ones. Although the physical mechanism about the seismic ionospheric anomalies is still unclear, a significant number of observational studies suggest that there is indeed a connection between the two phenomena. In general, the seismic ionospheric disturbance mainly includes the "earthquake precursor" effect of ionospheric TEC and plasma parameters.

There are currently two major types of methods for the measurement of seismic associated ionospheric anomalies: the ground-based stations and space-based satellites. The total electron content (TEC) derived from measurements of local ground-based GPS receivers was first employed by Liu et al. (2001) to study ionospheric electron density variations during the 1999 $M_W 7.6$ ChiChi earthquake and he found that the GPS TEC around the epicenter dramatically decrease in the afternoon period a day, 3 days, and 4 days before the earthquake. After that, Liu et al. (2004) further confirmed this pre-seismic precursor by conducting a statistical investigation of global ionospheric map (GIM) based on 20 $M_W \geq 6.0$ earthquakes during a period of 4 years from 1999-2002 in Taiwan., which demonstrates that the GPS TECs significantly decrease in the afternoon/evening period within 5 days prior to 16 of the 20 earthquakes. Following those, a number of related investigations were conducted by applying the GIM to study TEC anomalies before strong earthquakes with more reliable statistical methods. For instance, clear precursory positive anomalies of ionospheric total electron content (TEC) were found around the focal region prior to the 2011 Mw9.0 Tohoku-Oki earthquake

(Liu et al.,2011; Heki et al.,2011).. While, Kon et al. (2011) analyzed $M_w \geq$ 6.0 earthquakes which occurred in Japan from 1998 to 2010 by the superposed epoch analysis (SEA) method, and the positive TEC anomalies 1–5 days ahead were detected within 1000 km from the epicenters. It is also found that the TEC over the epicenter significantly enhances on a day before the 12 January 2010 M7 Haiti earthquake. The TECs of the two Mid-latitude dense strips on 35° N/60° S and those of Seismic-ionospheric anomalies in ionospheric TEC and plasma density of the epicenter/conjugate point reach their maximum values on a day before the earthquake, while the northern crest of equatorial ionization anomaly (EIA) moves poleward (Liu et al., 2011).

In most cases, however, the measurement of ground-based stations can be rather limited. Since there is a lack of extensive ground experiments to monitor geophysical and geochemical parameters in most areas. Thus, space-based satellite experiment with the vast coverage of the seismic areas of Earth can be regarded as a more effective way for measurements of seismic-ionospheric effects (Akhoondzadeh et al., 2010). The DEMETER (Detection of ElectroMagnetic Emissions Transmitted from Earthquake Regions) satellite data have already been applied to many studies. With the DEMETER data, a number of perturbations have been found before some strong earthquakes. Anomalies in the $O^+$ density, ion temperature, electric field, and ELF/VLF/ULF emissions around the epicenter region detected by DEMETER were considered to be highly associated with the 12 May 2008 M8.0 Wenchuan earthquake (Zhang et al. 2009a, b). A statistical investigation by Akhoondzadeh et al. (2010) made the simultaneous observations of positive and negative anomalies in both DEMETER and GPS data during 1-5 days before all studied earthquakes under weak and quiet geomagnetic conditions, which is highly regarded as pre-seismic precursor. Using more than 6 years observation data of DEMETER, Zhang et al. (2013) found that there are increases in the number of electron bursts events prior to the seismic activities; during the entire operation period of the DEMETER satellite, along with electron burst precipitation occurred before each strong earthquake with magnitude over 7.0.

Ionospheric electromagnetic perturbations were found by Zhima et al. (2012a, b) 4 days before the earthquake in the ELF/VLF frequency range. With the plasma data from DEMETER, Tao et al. (2017) found that both the electron density (Ne) and ion density (Ni) pronouncedly increased, the $O^+$ density increased and $H^+$ density decreased while the $He^+$ density remained relatively stable 2 days before the Java M7.7 earthquake in 2006.

Due to the importance and promising prospect of research about pre-seismic ionospheric anomalies, China Seismo-Electromagnetic Satellite (CSES) was launched on February 2, 2018 to monitor and study the seismic-ionospheric perturbations, and analyze the features of seismic-ionospheric perturbations. Several significant results were found during its first two years in orbit. With CSES data, Yan et al. (2018) studied 4 $M_w \geq 7.0$ earthquakes in 2018 and their results indicated unusual positive ionospheric perturbations in electron density, electron flux, VLF spectrogram, ion density and ion drift velocity 1-10 days before the studied earthquakes. It is also revealed by Song et al. (2020) that pre-seismic anomalies in electron density and total electron content (TEC) before 4 $M_w \geq 5.0$ earthquakes in 2018 by cross-validation analyzing the data from CSES, IRI-2016 model and total electron content (TEC) data from Center for Orbit Determination in Europe (CODE). In this paper, in order to analyze the features of seismic-ionospheric anomalies, but also to further verify the reliability of CSES scientific observation data, we investigated the seismic-ionospheric perturbations associated with the 14 July 2019 $M_w$7.2 Laiwui earthquake by cross-validation analyzing the GPS TEC data and data from different payloads of CSES (LAP and PAP). The basic information about seismic event and GPS satellite, CSES are briefly introduced in Section 2. The methodology and research results are presented in Section 3. In the end, discussion and conclusions of this study are implemented in Section 4.

## 2. Basic Information

2.1 Seismic Event Information

Indonesia is one of the most seismically active regions in the world, with comparatively much higher probability of seismic events occurrences. The frequent occurrence of earthquakes in this area provides an excellent chance and condition to study the phenomenon of seismic-ionospheric anomalies. Consequently, in this paper we took the Indonesian Laiwui earthquake occurred on 14 July 2019 as our research example. The magnitude $M_w$ 7.2 earthquake occurred in Indonesia Laiwui (-0.52°S, 128.17°E) with 10 km in depth at 09:10 UT (universal time) on July 14 2019. The radius of the Laiwui earthquake preparation zone estimated by the Dorbrovosky formula $\rho = 10^{0.43M}$ is about 1247.38 km. Figure 1 shows the location of the earthquake epicenter and preparation zone on the map.

2.2 GPS Satellite Data

The GPS satellites transmit two L-band signals at the frequencies of 1575.42 and 1227.60 MHz and offer an effective method for monitoring the ionosphere. The TEC is a measure of the total number of electrons that would be contained in a cylinder that extends up vertically above a given point on the Earth all the way through the ionosphere. The network of GPS receivers can be used to simultaneously and continuously monitor the TEC. (Liu et al. 2004)

To investigate the TEC variations, the GIM data provided by NASA Jet Propulsion Laboratory (JPL) were adopted to this study. The GIM is constructed into $5° \times 2.5°$ (Longitude, Latitude) grid with time resolution of 2 hour. GIM data are generated using data from 150 GPS sites of the IGS and other institutions. In our study, the TEC data based on the date and geographic location of Laiwui earthquake from 75 days before to 10 days after (30 April to 24 July) the main shock occurred.

2.3 China Seismo-Electromagnetic Satellite Data

The China Seismo-Electromagnetic Satellite (CSES), which is also named as

Zhang Heng-1 (ZH-1), was successfully launched on February 2, 2018. The CSES is the first space-based platform in China for both earthquake observation and geophysical field measurement, and it is a sun-synchronous satellite orbiting at a height of approximately 507 km with a descending node of 14:00 local time, an ascending node time of 02:00 LT with an inclination of $97.4°$. The distance between its neighboring tracks is 2650 km ($24°$ in longitude) in one day, while reduced to 530 km ($4°$ in longitude) in a revisit period of 5 days (Yan et al. 2018). The main objectives of this mission are to monitor the near-Earth space environment and investigate possible electromagnetic perturbations related to natural disasters and human activities. (Shen et al. 2018)

The scientific payload of the CSES is composed of several instruments that provide a nearly continuous survey of ionospheric plasma, waves, and energetic particles. In this study, the electron density and electron temperature data derived from LAP (Langmuir Probe), ion density ($He^+$, $O^+$) and ion temperature data derived from PAP (Plasma Analyzer Package) were applied to this research. Besides, also as shown in Figure 1 there were about 100 flight orbits (most are revisited orbits) above the earthquake region from one month before to 10 days after the earthquake (14 June to 24 July, 2019), which provided a significant amount of scientific observation data to our study.

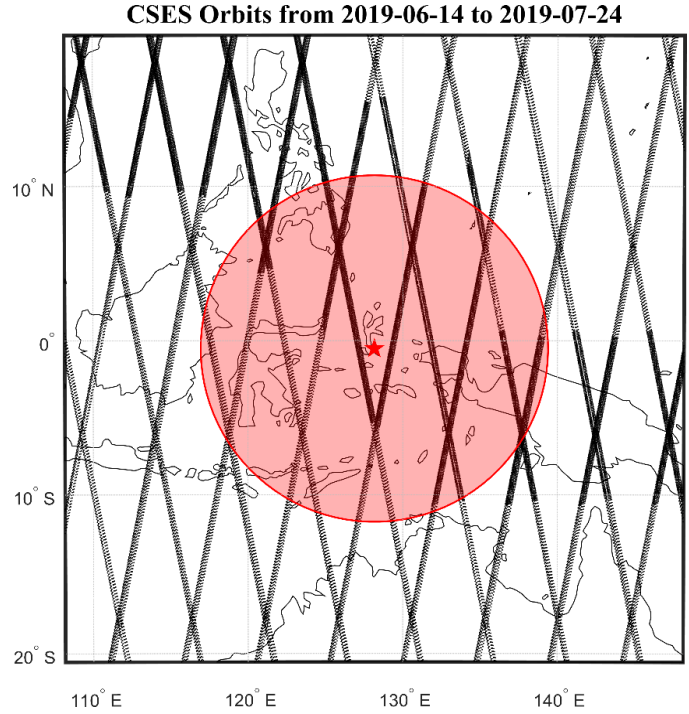

**Figure 1.** The red star and circle represent the epicenter and preparation zone of $M_w$ 7.2 Laiwui earthquake and CSES orbits from 2019-06-14 to 2019-07-24 are marked with black points.

## 3. Methodology and Research Results

3.1 TEC anomalies

The moving median and inter-quartile scope of data are used to shape the upper and lower bounds so that the seismic-anomalies could be separated from the background (Liu et al.,2004). In addition, to calculate the statistical parameters, the length of the period was selected as about 55 days in order to avoid affects by the seasonal variations. The upper and lower bound of the mentioned range can be calculated using the following equations (1)-(4):

$$TEC_{\text{UB}} = TEC_{M30} + k \cdot TEC_{IQR} \qquad (1)$$

$$TEC_{\text{LB}} = TEC_{M30} - k \cdot TEC_{IQR} \qquad (2)$$

$$\Delta TEC = \frac{(TEC_{obs} - TEC_{M30})}{TEC_{IQR}} \qquad (3)$$

$$p = \pm[(\varDelta TEC - k)/k] \cdot 100\% \qquad (4)$$

where $TEC_{M30}$, $TEC_{IQR}$, $TEC_{UB}$, $TEC_{LB}$, $TEC_{obs}$, $\varDelta TEC$ and $k$ are the 30-day TEC moving median value, TEC inter-quartile range, TEC upper bound, TEC lower bound, TEC observed value, differential of TEC and threshold of the anomaly, respectively. Here we set the $k = 2.0$ considering the magnitude of the main shock to select the anomalies interval. Over and above, while the absolute value of $\varDelta TEC$ is larger than the $k$ value ($|\varDelta TEC| \geq k$), the behavior of the pertinent TEC value will be noted as anomalous.

Basically, we check the variations of the geomagnetic data including Dst, Kp index and solar flux F10.7 index variation during 30 May to 24 July 2019, i.e., 45 days before to 10 days after the $M_w 7.2$ Laiwui earthquake. Furthermore, a harsh condition (Dst > −30 nT, Kp < 3 and F10.7 < 100 sfu) is adopted to distinguish pre-seismic ionospheric phenomena triggered by solar activities. Fig.2 shows that geomagnetic and solar activities are relatively weak and quiet during that period except a magnetic storm occurred on 10 July 2019, which is marked by red arrows and dashed elliptic. By a linear interpolation of 4 data points which is adjacent the epicenter (0.52°S, 128.17°E), we calculate the TEC above the epicenter. In consideration of the resolutions of the latitude and longitude( 2.5° in latitude and 5° in longitude) in GIM TEC , the ranges of 125° − 130° E and 0−2.5° S are selected as the data points center. Seen from Fig.2d, it represents the $\varDelta TEC$ values between 30 May 2019 and 24 July 2019 using Eq.(3).

In addition, anomalous TEC times are picked searched with $|\varDelta TEC|$>2.0, Dst<−30nT, Kp<3 and F10.7<100 sfu. The anomalies are found in 8 days (6 July) before the earthquake, 3 days (11 July) before the earthquake, 1 day (13 July) before the earthquake and 6 days after the earthquake (the main shock onset marked with a red star) illustrated in Fig.2d. Likewise, the anomalies can be positive as well as negative which is consistent with previous researches ( Akhoondzadeh et al., 2010; Pulinets and Davidenko, 2014; Pulinets et al., 2003, 2015). We conclude there are increases of TEC anomaly during the interval of 06:00-08:00 UT 6 July (15:00-17:00) LT, +22.28%

enhances, 08:00-10:00 UT 11 July (17:00-19:00) LT, +13.85% enhances, 06:00-08:00 13 July (15:00-17:00) LT, +24.52% enhances and decrease of TEC anomaly during the interval 16:00-18:00 UT 20 July (01:00-03:00) LT, 22.75% decreases.

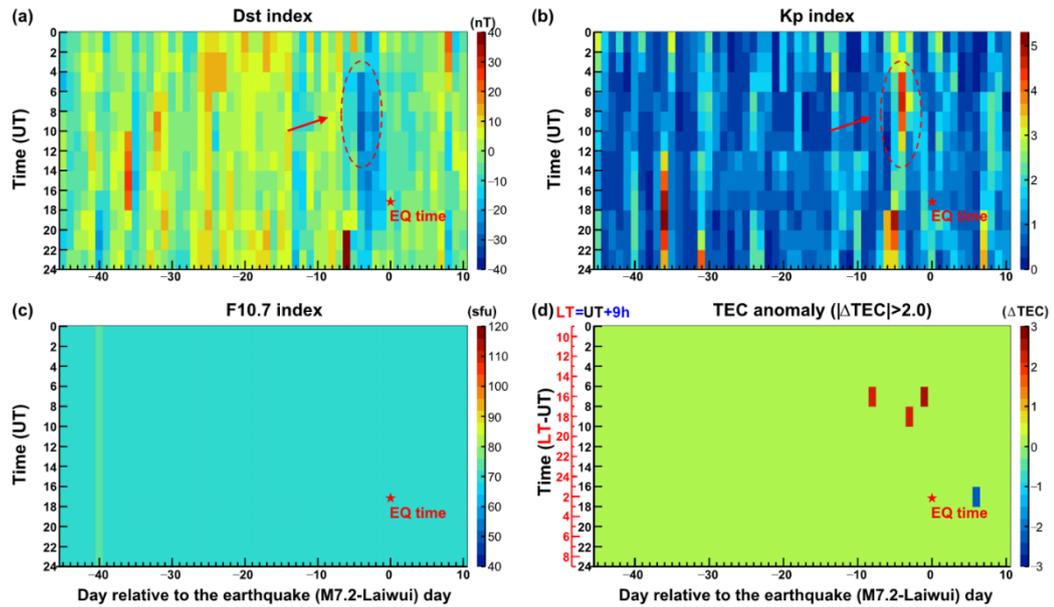

**Figure 2.** TEC anomaly analysis for the Laiwui earthquake (14 July 2019) from 30 May 2019 (45 days before the earthquake) to 24 July 2019 (10 days after the earthquake). The earthquake time is represented by a red star. The x axis represents the day relative to the earthquake day. The y axis represents the UT (LT = UT + 9 h). (a) Dst geomagnetic index, the magnetic storm occurred on 10 July is marked with the red arrow and dashed elliptic. (b) Kp geomagnetic index, the magnetic storm occurred on 10 July is marked with the red arrow and dashed elliptic. (c) Solar radio flux F10.7 index. (d) TEC anomalies detected under the following conditions: Dst > −30 nT, Kp < 3, F10.7 < 100 sfu and $|\Delta TEC| > 2.0$. Here 1 TECU = $10^{16}$ electrons/$m^2$.

3.2 Geographical anomalies on TEC with latitude-longitude-time (LLT) maps

With regarding to aforesaid four anomalous intervals, there is a geographical investigation to check whether the GIM TEC concurrently disturbs in that earthquake locality. Respectively, every GIM map consists of 5040 (70 * 72) grids and covers ±87.5° N latitude and ±180° E longitude ranges with spatial resolutions of 2.5° in latitude and 5° in longitude.

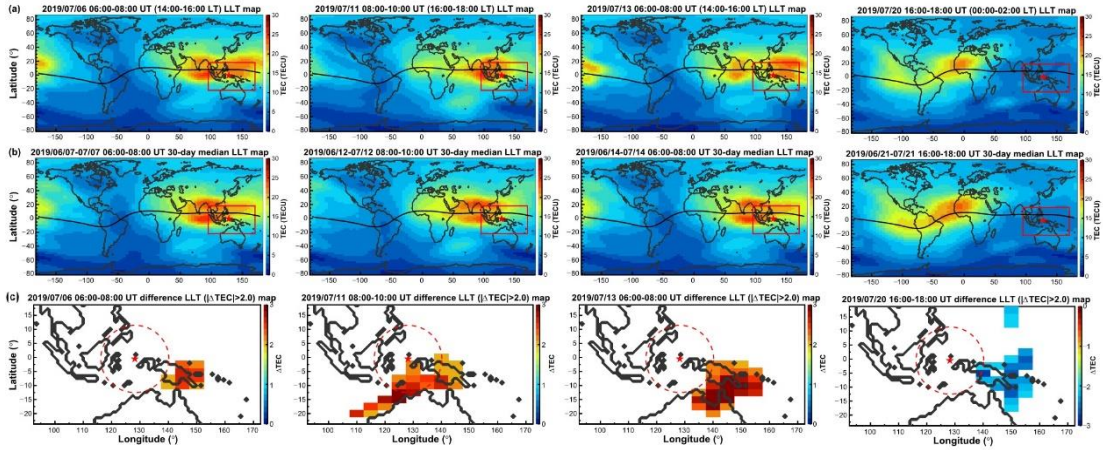

**Figure 3**. The GIM latitude-longitude-time (LLT) maps observed during the interval of 06:00-08:00 UT 6 July, 08:00-10:00 UT 11 July, 06:00-08:00 13 July before the 14 July 2019 $M_w$7.2 Laiwui earthquake and 16:00-18:00 UT 20 July after the 14 July 2019 $M_w$7.2 Laiwui earthquake. The GIM LLT maps during the fixed period of 06:00–08:00 UT 6 July 2019 (1st column), 08:00–10:00 UT 11 July 2019 (2nd column), 06:00–08:00 UT 13 July 2019 (3rd column) and 16:00-18:00 UT 20 July 2019 (4th column). Panels of row (a) are the observed values on 8 days before the earthquake (6 July 2019), 3 days before the earthquake (11 July 2019), a day before the earthquake (13 July 2019) and 6 days after the earthquake (20 July 2019), while row (b) shows the median values of the period of days 1–30 before each anomalous interval. The red squares in rows (a, b) indicate the regions of interest around the earthquake, in range of 22°S–18°N latitude and 95°-170°E longitude. Panels of row (c) denote the extreme differences ($|\Delta TEC| > 2.0$) of the 30-day period that appeared on 8 days before the earthquake (6 July 2019), 3 days before the earthquake (11 July 2019), a day before the earthquake (13 July 2019) and 6 days after the earthquake (20 July 2019) with the regions of interest around the earthquake. The color denotes the difference value of the TEC from the relevant median value. The red dashed circles with the radius $\rho$=1247.38km represent the earthquake preparation area of the lithosphere.

As seen in Fig.3a, the row of GIM TECs LLT map is for each anomalous interval (06:00-08:00 UT 6 July, 08:00-10:00 UT 11 July, 06:00-08:00 UT 13 July and 16:00-

18:00 UT 20 July). The median of each grids on GIM TECs in each anomalous above-mentioned interval during 1-30 days before each anomalous interval is shown in Fig.3b. Fig.3d shows ultimate percentage of TEC ($|\Delta TEC| \geq 2.0$) between the observed GIM TEC and the associated 30-day median at four anomalous intervals occurred on 06:00-08:00 UT 6 July (1st column)., 08:00-10:00 UT 11 July (2nd column), 06:00-08:00 13 July (3rd column) and 16:00-18:00 UT 20 July (4th column), respectively. Generally, the 30-day median is on behalf of the undisturbed background, whereas positive percentage of TEC represents the enhancement of GIM TECs but negative percentage of TEC represents the decrease of GIM TECs.

As shown in Fig.3c, the GIM TECs around Laiwui earthquake epicenter dramatically enhance by ~1.14-31.03% in the interval of 06:00-08:00 UT 6 July (15:00-17:00 LT), ~0.75-56.98% in the interval of 08:00-10:00 UT 11 July (17:00-19:00 LT), ~2.75-66.68% in the interval of 06:00-08:00 13 July (15:00-17:00 LT) and decrease ~2.70~41.38% in the interval of 16:00-18:00 UT 20 July (01:00-03:00 LT).

The sequence of GIM for four corresponding global fixed local times is examined in exchange for eliminating the local time and/or EIA effects. As shown in Fig. 3 compared with the TEC enhancements at four different universal times in Fig. 4, the corresponding extreme enhancements in the GIM TECs at global fixed local times are also chiefly positioned around the forthcoming epicenter and EIA region. Accordingly, the geographical anomalies simultaneously and remarkably appear in the four anomalous intervals around the epicenter of the Laiwui earthquake. Concretely, the GIM TECs around Laiwui earthquake epicenter dramatically enhance by ~6.98-65.31% in the interval of 06:00-08:00 UT 6 July (15:00-17:00 LT), ~0.45-19.28% in the interval of 08:00-10:00 UT 11 July (17:00-19:00 LT), ~10.00-62.16% in the interval of 06:00-08:00 13 July (15:00-17:00 LT) and decrease ~5.88-23.53% in the interval of 16:00-18:00 UT 20 July (01:00-03:00 LT).

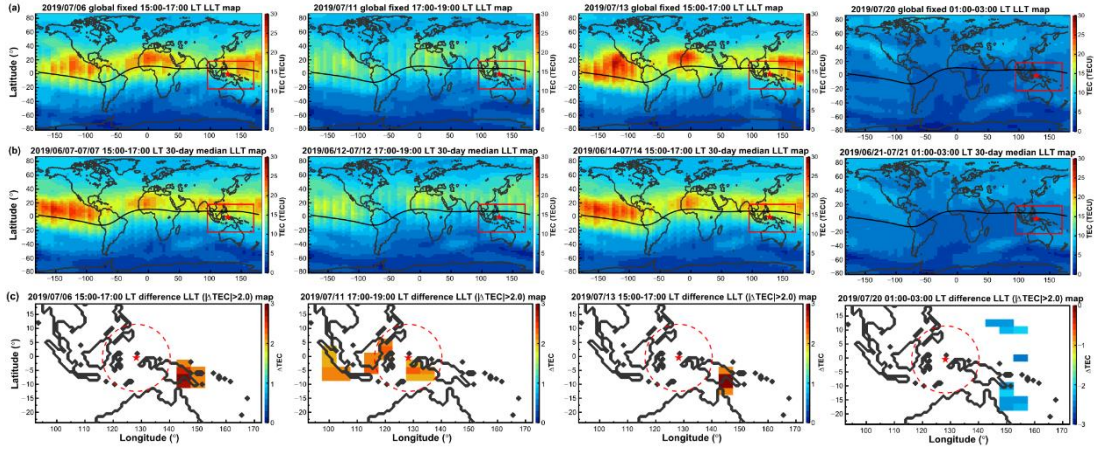

**Figure 4**. The GIM LLT maps observed during the global fixed intervals of 15:00-17:00 LT 6 July, 17:00-19:00 LT 11 July, 15:00-17:00 LT 13 July before the earthquake and 01:00-03:00 LT 20 July after the earthquake. The GIM LLT maps during four global fixed local times: (1st column) 15:00-17:00 LT 6 July 2019, (2nd column) 17:00-19:00 LT 11 July 2019, (3rd column) 15:00-17:00 LT 13 July 2019 and (4th column) 01:00-03:00 LT 20 July 2019, respectively. Panels of row (a) are the observed values on 8 days before the earthquake (6 July 2019), 3 days before the earthquake (11 July 2019), a day before the earthquake (13 July 2019), and 6 days after the earthquake (20 July 2019), while row (b) shows the median values of the period of days 1–30 before each anomalous interval. The red squares in rows (a, b) indicate the regions of interest around the earthquake, in range of 22°S–18°N latitude and 95°-170°E longitude. Panels of row (c) denote the extreme differences ($|\Delta TEC| > 2.0$) of the 30-day period that appeared on 8 days before the earthquake (6 July 2019), 3 days before the earthquake (11 July 2019), a day before the earthquake (13 July 2019) and 6 days after the earthquake (20 July 2019) with the regions of interest around the earthquake. The color denotes the difference value of the TEC from the relevant median value. The red dashed circles with the radius $\rho$= 1247.38km represent the earthquake preparation zone.

### 3.3 Plasma parameters perturbations

In this study, GIM TEC anomalies derived from GPS satellites 45 days before to 10 days after the earthquake have already been analyzed. To further confirm the observed TEC anomalies, a cross-valid examination is conducted with the application of the observation data from CSES.

As introduced in Section. 2.3, the data recorded by payloads LAP and PAP on CSES are adopted to study the ionospheric plasma parameters perturbations above the earthquake preparation zone during the period of 30 days before (14 June 2019) to 5 days after (19 July 2019) the Laiwui earthquake. We examine the percentage deviation of the plasma parameters recorded by CSES within 30 days before and 10 days after the earthquake via moving mean method. Deviation of the plasma parameters can be calculated using the following formulas:

$$dN = \frac{N_{obs} - N_{mean}}{N_{mean}} \times 100\% \tag{4}$$

$$dT = \frac{T_{obs} - T_{mean}}{T_{mean}} \times 100\% \tag{5}$$

$N_{obs}$ and $T_{obs}$ are the CSES observed values for each plasma parameter, while $N_{mean}$ and $T_{mean}$ are perceived as the background values, which are the corresponding moving means from previous 30 days orbit data (data cell is sampled by 2° in latitude and 4° in longitude). Indeed, unusual perturbations in different ionospheric plasma parameters are detected prior to 14 July Laiwui earthquake. Figure 5 (a) to (e) displays the percentage deviation of electron density ($Ne$), electron temperature ($T_e$), $O^+$ density ($N_{O^+}$), ion temperature ($T_i$) and $He^+$ density ($N_{He^+}$) respectively. However, due to the measurement limitation of the PAP instrument, there is little valid data for $H^+$ density above the earthquake area and the measurement of $He^+$ density is also not persistent for some certain orbits.

From the TEC anomalies analysis, the TEC anomalies were detected on 8 days (6 July), 3 days (11 July) and 1 day (13 July) prior to the earthquake. A further cross-valid analysis is conducted during these periods. As shown in Figure 6(a), the electron density increased significantly on day 4 (10 July) and day 2 (12 July) before the earthquake, the maximum value increased by approximately 135.32% and 115.69% respectively when it approached the epicenter. While, on day 3 (11 July) and day1 (13 July) before the earthquake, the maximum only increased by about 16.80% and 39.58%. Similarly, as shown in Figure 6(c) the main component $O^+$ density also increased dramatically on day 4 (10 July) and day 2 (12 July) before the seismic event when it approached the epicenter, the maximum value increased by 160.10% and 153.74% respectively. While, the $O^+$ density remained relatively stable 3 days before and 1 day before with a slight increase about 10.63% and 21.73%. Although the observation data of $He^+$ density is not persistent for some orbits, as shown in Figure 6(e), the variation tendency of $He^+$ density can still be observed from the data recorded by CSES. $He^+$ density profoundly increased when flying above the epicenter on 4 days before the earthquake with maximum increased by 154.76%. The electron temperature and the ion temperature, however, remained relatively stable during the observation period, with a comparatively slight increase no more than 60% for all orbits.

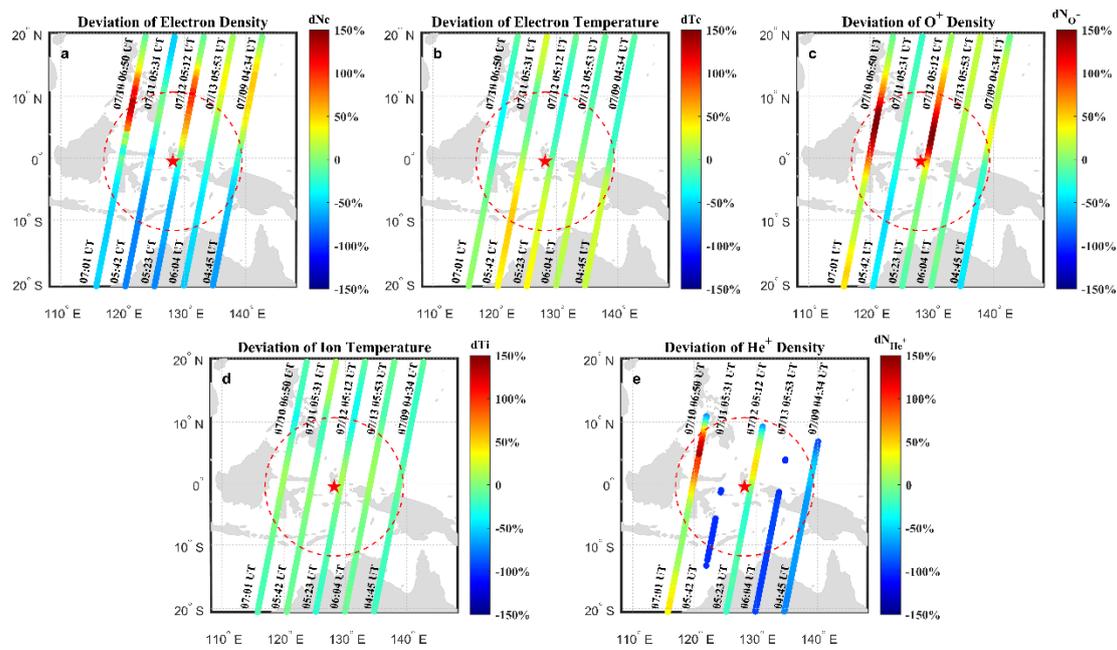

**Figure 5** Deviation percentage of plasma parameters from 9 July to 13 July. The red stars represent the epicenter of the earthquake and the red dashed circles represent the preparation zone ($\rho$ =1247.38 km), the precise moments (universal time) of flying above this area are marked at the beginning and end of each orbit.

Also, Figure 6(a) to (e) displays the percentage deviation of the same plasma parameters, while the observation period is from 4 July to 8 July. Since a magnetic storm occurred on 8 July, as shown in Figure 2, it is difficult to distinguish whether the anomalies on 8 July is caused by the magnetic storm or the earthquake. As displayed in Figure 6(a), the electron density dramatically increased on 5 July and 7 July when approaching the epicenter with a maximum increase by approximately 129.29% and 151.17% respectively. While the electron density on 4 July and 6 July remained relatively stable, with the greatest deviation percentage no more than 80%. $O^+$ density increased significantly by the order of 112.61% and 197.77% on 5 July and 7 July, while the adjacent orbits remained relatively stable. Besides, $He^+$ density also increased simultaneously with the $O^+$ density on 7 July, with a maximum increase by 186.29% Furthermore, the variation of the electron and ion temperature still remained relatively stable (deviation no more than 50%) without significant perturbations during the observation period.

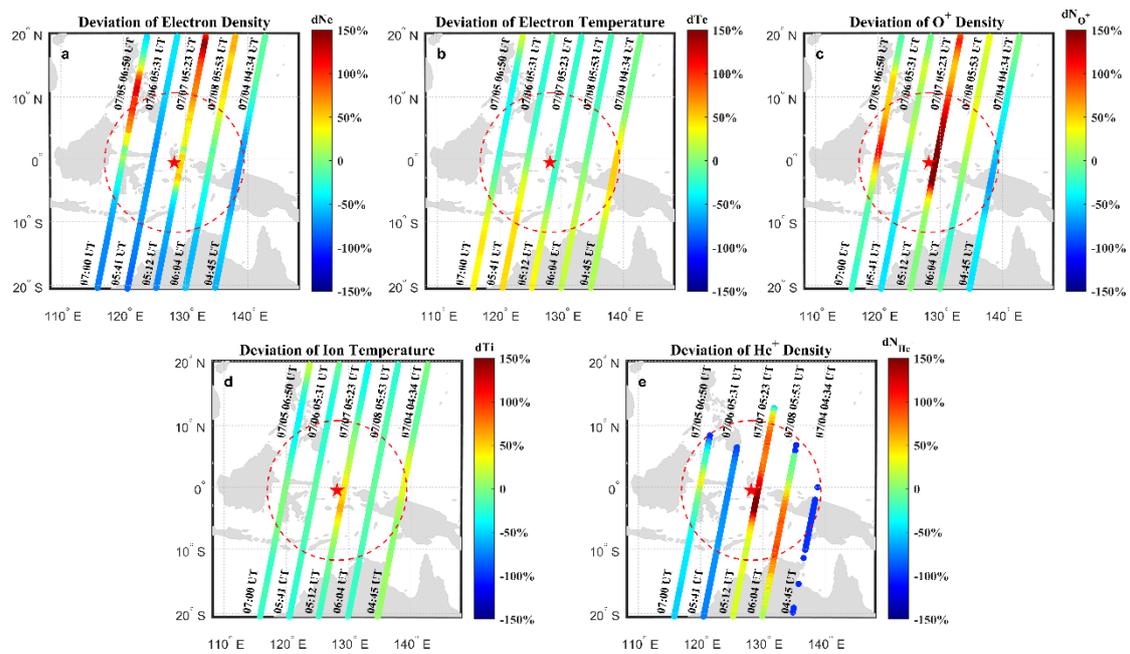

**Figure 6** Deviation percentage of plasma parameters from 4 July to 8 July. The red stars represent the epicenter of the earthquake and the red dashed circles represent the preparation zone ($\rho$ =1247.38 km), the precise moments (universal time) of flying above this area are marked at the beginning and end of each orbit.

To further verify the unusual variations of the in-situ parameters, the electron density (Ne) and electron temperature (Te) data of the abnormal orbits were extracted to make comparison with their corresponding revisited orbits. Figure 7(a) and (b) represent the variation of electron density of 12 July and 10 July along with their corresponding revisited orbits. It can be clearly observed that the electron density increased significantly within $0° - 20°$N, with a peak value reaching $9.21\times 10^{10} N_e/m^3$, $9.59\times 10^{10} N_e/m^3$ on 12 July and 10 July respectively. Although due to the equator ionospheric anomaly (EIA), the electron density all increased near the magnetic equator ($7.6°$N), it can still be observed that the abnormal orbits did show unusual positive anomalies, compared with their corresponding revisited orbits. Figure 7(c) and (d) demonstrate the change of electron temperature of the same orbits, and it is presented in the results that Te is inversely proportional to Ne. This phenomenon is caused by the cooling process of electrons, of which the rate is proportional to the square of Ne and consistent with the basic ionospheric theory concerning the relationship between these two parameters. ((Bilitza, 1975; Bilitza and Hoegy, 1990; Kakinami et al., 2011; Song et al., 2020)

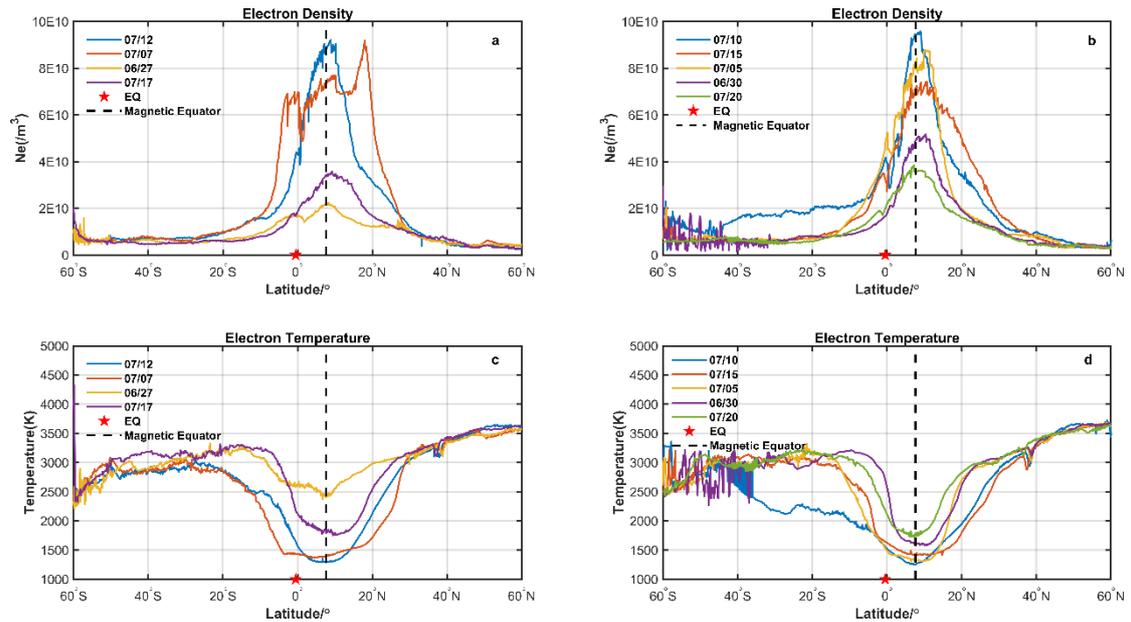

**Figure 7** The variations in Ne and Te from the abnormal orbits, and their revisiting orbits along latitude. Red stars represent earthquake epicenters, and the black dotted line in each subfigure represents the magnetic equator.

Furthermore, it also should be noted that CSES usually flew above the earthquake preparation zone twice a day, with a descending orbit and an ascending orbit respectively. The obvious disturbances were observed by CSES when it was flying above the region with descending orbits. However, we do not find any similar variations in plasma parameters before the occurrence of earthquake during ascending orbits at about 02:00 LT (17:00 UT).

## 4. Discussion and Conclusion

In this study, seismic-ionospheric anomalies associated with the $M_w$7.2 Laiwui earthquake were comprehensively investigated. Regarding the temporal distribution, from the observation of the GPS satellite, there were three significant positive perturbations in TEC 1, 3, 8 days prior to the earthquake and a negative disturbance six days after the earthquake under the relatively quiet geomagnetic conditions. In respect of the spatial distribution, considering the local time and EIA effects, the spatial distribution and the signs of the anomalies agree well with those of the TEC anomalies (Figure. 2d) on each day. To make the results more convincing, more extended types of investigations were carried out, such as in-situ investigations of plasma parameters variations using the data from LAP, PAP of CSES. All the results of these investigations indicate the seismic-ionospheric anomalies prior to the Laiwui earthquake. It should be noted that the unusual TEC perturbations on 7 July may also be associated with the $M_w$6.9 earthquake, which occurred on 7 July in Kota Ternate (0.513°N, 126.19°E) near the epicenter.

Nevertheless, further discussions are required for some of the results. A cross-valid analysis of TEC anomalies was conducted using the LAP and PAP data of CSES, and the results are indeed consistent with the former one. There were great enhancements in plasma parameters (electron density, ion density, etc.) prior to the earthquake, however, there was also a difference between the two results. For example, the most significant anomalies of the electron density were observed on day 4 (10 July) and day 2 (12 July) before the earthquake, while the TEC anomalies were exactly one day after the CSES observations. Besides, a negative anomaly in TEC was also detected 6 days after the earthquake, but similar disturbances were not detected by CSES. There might be several reasons accounting for these discrepancies, on the one hand, as shown in

Figure 4b, the Kp index increased at 4:00-10:00 UT on 10 July, while the electron density also increased simultaneously during the same period, so the disturbances on 10 July may be related to the geomagnetic activity. On the other hand, the discrepancies may be mainly attributed to the difference of the two datasets. To be specific, CSES is a spacecraft exploring the topside ionosphere at an altitude about 507 km with in-situ observations, while the GPS-TEC is calculated under the assumption of the ionospheric single layer.

Besides, some of the TEC anomalies were observed outside the preparation zone., as shown in Figure 3 and Figure 4. This can be attributed to lithosphere-atmosphere-ionosphere coupling (LAIC) process, the Dobrovolsky formula is an ideal equation without the consideration of the LAIC process. The earthquake-related anomalies induced by the LAIC mechanism works complicatedly, whose wave channels mainly is composed of the acoustic-gravity wave (AGW) propagation, electromagnetic emission (EME) and geochemical channel (Pulinets and Boyarchuk, 2004; Kamogawa, 2006; Hayakawa, 2006; Kuo et al., 2014; Pulinets and Davidenko, 2014). Therefore, the seismic-ionospheric anomalies may propagate to further distance.

As for the observation discrepancies in different CSES orbits (ascending and descending ones), this may be attributed to the ionospheric daily variation. The descending orbits of the CSES usually flew above the earthquake region at about 15:00 LT (06:00 UT), while the ascending orbits passed there at about 02:00 LT (17:00UT). At daytime, the ionosphere received much more radiation from the sun resulting in more ionized particles, which significantly increases the density of the ionospheric electrons and ions. While the densities of the electrons and ions are much lower at night time. Due to the relatively lower electron and ion densities, the variations of the ionospheric plasma parameters will be much more difficult to be detected by CSES. Thus, the plasma parameters observed by CSES remained relatively stable during night time prior to the earthquake. Also, perturbations in electron density occurred more often than those of electron temperature from the observation results of CSES, which illustrate that electron density is much more sensitive to seismic activity than electron temperature, this is also consistent with the statical investigation conducted by Liu et al. (2014). Besides, it also should be noted that the PAP instrument of CSES is slightly contaminated, with lower absolute value in observation data, therefore the data of PAP can only be applied to the relative deviation analysis.

In conclusion, during these periods, the measurements of GPS and CSES yield similar tendencies, the temporal and spatial anomalies of the TEC and ionospheric plasma perturbations detected by CSES over the epicenter did indicate significant positive seismic-ionospheric anomalies. Based on the results presented, we can also safely draw the conclusion that CSES data are reliable for the study of seismic events. Also, the localization and synchronization of the longtime anomalies around the occurrence of earthquake suggest that these perturbations are highly associated with $M_w$7.2 Laiwui earthquake, but further investigations are required in the future to obtain a more accurate knowledge of the perturbation process.

**Acknowledgement**

The authors would like to thank Dr. Rui Yan at the Institute of Crustal Dynamics and Mr. Hongyi Fu at Chengdu University of Technology for helpful discussions. The authors also would like to thank Mr. Hengxin Lu and Mr. Dapeng Liu at CSES ground application center for the CSES data services.